\documentclass[twocolumn,english,prb,aps,superscriptaddress]{revtex4-1}
\usepackage[T1]{fontenc}
\usepackage[latin9]{inputenc}
\usepackage{units}
\usepackage{bm}
\usepackage{amsmath}
\usepackage{amssymb}
\usepackage{graphicx}
\usepackage{esint}

\makeatletter

\@ifundefined{textcolor}{}
{%
\definecolor{BLACK}{gray}{0}
\definecolor{WHITE}{gray}{1}
\definecolor{RED}{rgb}{1,0,0}
\definecolor{GREEN}{rgb}{0,1,0}
\definecolor{BLUE}{rgb}{0,0,1}
\definecolor{CYAN}{cmyk}{1,0,0,0}
\definecolor{MAGENTA}{cmyk}{0,1,0,0}
\definecolor{YELLOW}{cmyk}{0,0,1,0}
}

\usepackage{bm}

\@ifundefined{textcolor}{}{%
 \definecolor{BLACK}{gray}{0}
 \definecolor{WHITE}{gray}{1}
 \definecolor{RED}{rgb}{1,0,0}
 \definecolor{GREEN}{rgb}{0,1,0}
 \definecolor{BLUE}{rgb}{0,0,1}
 \definecolor{CYAN}{cmyk}{1,0,0,0}
 \definecolor{MAGENTA}{cmyk}{0,1,0,0}
 \definecolor{YELLOW}{cmyk}{0,0,1,0}
 }

\usepackage{xcolor}

\makeatother

\usepackage{babel}
\begin{document}

\title{$8\pi$-periodic dissipationless ac Josephson effect on a quantum
spin-Hall edge via a Quantum magnetic impurity}

\author{Hoi-Yin Hui}

\altaffiliation{Current Affiliation: Department of Physics, Virginia Tech, Blacksburg, Virginia 24061, USA}

\selectlanguage{english}%

\affiliation{Department of Physics, Condensed Matter Theory Center and Joint Quantum
Institute, University of Maryland, College Park, Maryland 20742-4111,
USA}

\author{Jay D. Sau}

\affiliation{Department of Physics, Condensed Matter Theory Center and Joint Quantum
Institute, University of Maryland, College Park, Maryland 20742-4111,
USA}

\date{\today}
\begin{abstract}
Time-reversal invariance places strong constraints on the properties
of the quantum spin Hall edge. One such restriction is the inevitability
of dissipation in a Josephson junction between two superconductors
formed on such an edge without the presence of interaction. Interactions
and spin-conservation breaking are key ingredients for the realization
of dissipationless ac Josephson effect on such quantum spin Hall edges.
We present a simple quantum impurity model that allows to create a
dissipationless fractional Josephson effect on a quantum spin Hall
edge. We then use this model to substantiate a general argument that
shows that any such non-dissipative Josephson effect must necessarily
be $8\pi$-periodic. 
\end{abstract}

\pacs{PACS}

\maketitle

\section{Introduction}

The Josephson effect \cite{Josephson1964,Kulik2010,Golubov2004},
which was originally a direct manifestation of macroscopic quantum
coherence in superconductors, has turned out to be one of the most
reliable ways of diagnosing the topological properties of a junction.
Topological superconductors (TSCs) supporting Majorana modes have
been shown to demonstrate a $4\pi$-periodic Josephson effect \cite{Kitaev2001,Kwon2004,Fu2009,Tanaka2009,Lutchyn2010,Oreg2010,meyer},
which is doubled period compared to the conventional Josephson effect.
This phenomenon, which is known as the fractional Josephson effect,
has been observed in quite a few devices~\cite{rokhinson,Woerkom2016}
including the quantum spin Hall edge~\cite{molenkamp1,molenkamp2}.
At first this is quite counter-intuitive given that the Hamiltonian
itself is $2\pi$-periodic. The fractional Josephson effect in this
case arises because the topological property of the superconductor
forces the local fermion parity (FP) of the junction to change with
each rotation of the phase by $2\pi$. If one assumes contact with
a bath that equilibrates the system to the ground state of the appropriate
FP \cite{Beenakker2013} then the topological nature of the superconductor
is precisely reflected in the fractional Josephson effect. Interestingly,
the addition of interaction can often modify the topological classification
qualitatively \cite{Fidkowski2010,Tang2012}. In fact, some of the
states such as parafermion states in superconductors \cite{Mong2014}
are already known to be characterized by exotic Josephson effects.

While several topological superconducting phases involving interactions
have been proposed \cite{Mong2014,Oreg2014,Gaidamauskas2014}, not
many of them are within experimental reach. On the other hand, an
interesting $8\pi$-periodic Josephson effect, which relies on the
combination of interaction and topology that has recently been proposed~\cite{Zhang2014}
certainly appears to be within the realm of experimental possibility.
Ideally, one could just obtain this effect by studying a Josephson
junction (JJ) on an interacting two dimensional topological insulator
(TI) edge~\cite{Zhang2014}.

\begin{figure}
\begin{centering}
\includegraphics[width=1\columnwidth]{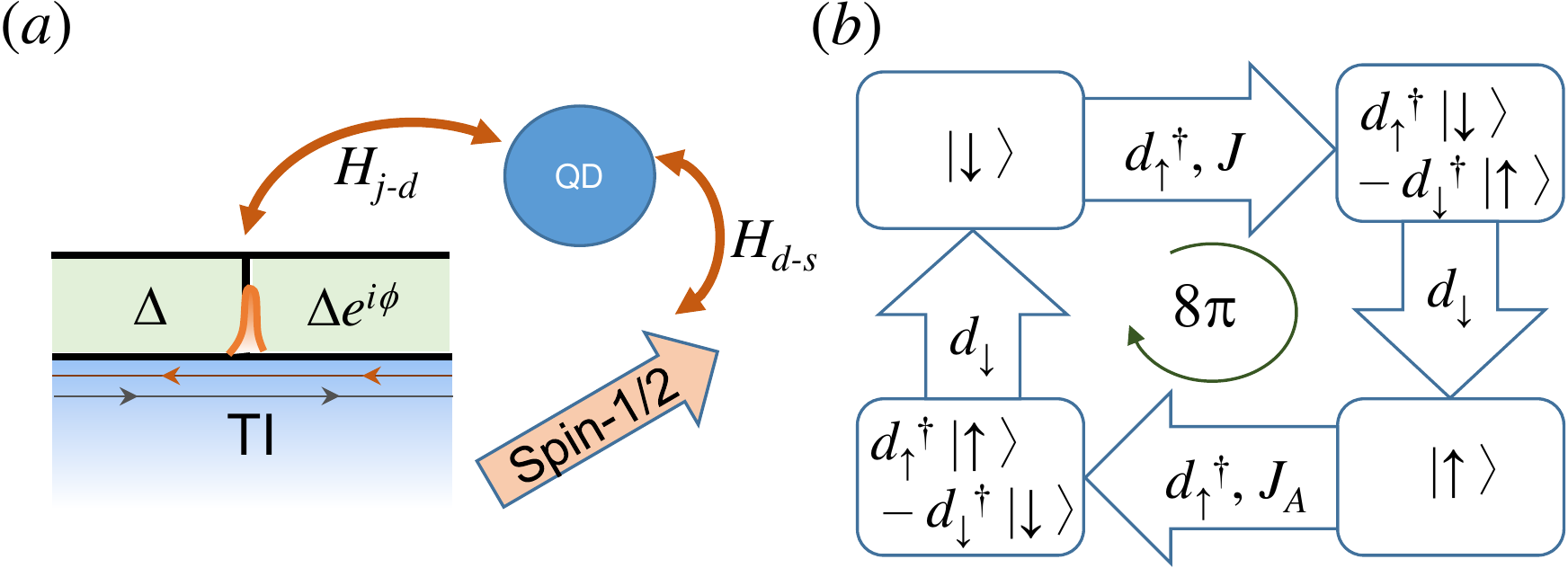} 
\par\end{centering}

\caption{(a) The system with a short Josephson junction whose Andreev state
is tunnel-coupled to a quantum dot. The quantum dot is coupled to
another localized spin. {[}Eq.~(\ref{eq:Hdotspin}){]} (b) Schematic
diagram showing the $8\pi$ cycle of states. $|\uparrow,\downarrow\rangle$
represent the states of the spin and $d_{\uparrow,\downarrow}^{\dagger}$
represents the electron in the quantum dot. $J,J_{A}$ represents
the spin symmetric and asymmetric exchange interactions inside the
dot respectively. In each $2\pi$ cycle of phase shown by the bold
arrow a spin up electron is pumped into or a spin down hole leaves
the edge.\label{fig:QDsys}}
\end{figure}

However, as we discuss in this paper, the $8\pi$ Josephson effect
turns out to be the only possible non-dissipative Josephson effect
that a quantum spin Hall edge can support. To understand what we mean
by non-dissipative Josephson effect consider a finite voltage biased
Josephson junction as is used to study the ac Josephson effect. The
ideal response of such a Josephson junction to a voltage bias is to
create an ac current that can be measured as radiation. However, at
large finite voltages, a typical Josephson junction dissipates part
of its energy through processes that generate quasiparticles in the
bulk~\cite{Averin1995}. The dissipation in turn leads to a dc current
in addition to the ac current, which is parametrized by the shunt
resistance of the junction. In this paper, we will be interested in
understanding the conditions under which such effect shunt resistances
can be avoided in TI junctions.

As we will review in more detail later, even though a non-interacting
TI edge with a ferromagnetic insulator (FI) is predicted to have a
$4\pi$-periodic fractional Josephson effect \cite{Fu2009,Tanaka2009},
removing the FI qualitatively modifies this effect. In the absence
of a relaxation mechanism or for a short junction, the JJ on the TI
edge has a $2\pi$ periodicity characteristic of conventional Josephson
junctions. The dissipation, which is observable as a parallel shunt
resistance across the junction \cite{Averin1995}, arises here from
ejection of quasiparticles into the conduction band. The addition
of a relaxation mechanism also leads to a dissipative but $4\pi$-periodic
fractional Josephson effect \cite{Beenakker2013}. Thus, one can say
that the Josephson effect on a non-interacting TI edge with TRS is
always dissipative and fundamentally accompanied by a shunt resistance.
As shown by Zhang and Kane~\cite{Zhang2014}, the addition of interaction
qualitatively changes this story and introduces a topologically protected
$8\pi$-periodic fractional Josephson effect, which is non-dissipative
(i.e. free of the shunt resistance). It is worth noting that this
effect, unlike the $4\pi$, $8\pi$ and $12\pi$-periodic Josephson
effect that can arise from fine-tuning in conventional systems \cite{Sothmann2013},
is indeed topologically protected in the sense that it is completely
robust to all perturbations of the Hamiltonian that preserve time-reversal
symmetry.

In this paper, we study the effect of a strongly interacting quantum
dot (QD) in a quantum spin-Hall Josephson junction. By considering
a simple model of such a QD that acts like a spin coupled to Andreev
bound states (ABS)~\cite{nazarov}, we show that such a junction
would show an $8\pi$-periodic fractional Josephson effect, in contrast
to the $4\pi$-periodicity expected from time-reversal breaking topological
junction. We then argue that the generic low voltage periodicity of
the quantum spin-Hall Josephson junction is $8\pi$ as opposed to
the $4\pi$ periodicity for time-reversal breaking topological junctions.
In fact, while higher frequencies could lead to $2\pi$ or $4\pi$-periodic
Josephson effects, such ac Josephson effects are necessarily dissipative
i.e. accompanied by a finite dc current.

\section{Junction with a quantum dot}

While the non-dissipative $8\pi$-periodic
Josephson effect is generic, we start by demonstrating it's origin
through the simple device shown in {[}Fig.~\ref{fig:QDsys}(a){]}.
This model incorporates the key ingredients for a non-dissipative
Josephson effect - namely, a topological quantum spin-Hall edge, spin-conservation
breaking and interaction. The device in {[}Fig.~\ref{fig:QDsys}(a){]}
consists of a JJ on a QSH edge laterally coupled to a strongly-interacting
multi-orbital quantum dot. The interaction is chosen to be large enough
so that the quantum dot admits at most two electrons. Furthermore,
the Hunds coupling is also assumed to be strong so that one of the
levels is always occupied by one electron, which can thus be considered
to be a local moment with spin ${\bf {S}}$. The resulting spin $-1/2$
is exchanged coupled to the dot electron with spin ${\bf {s}}$ via
the Hunds coupling through a Hamiltonian 
\begin{align}
 & H_{d-s}=J\bm{S}\cdot\bm{s}+J_{A}\left(S^{+}s^{+}+S^{-}s^{-}\right)+\varepsilon_{d}\left(d_{\uparrow}^{\dagger}d_{\uparrow}+d_{\downarrow}^{\dagger}d_{\downarrow}\right),\label{eq:Hdotspin}
\end{align}
 where the spin of the dot electron $d_{\sigma=\uparrow,\downarrow}^{\dagger}$
can be written as $\bm{s}=\frac{1}{2}\sum_{\alpha\beta}d_{\alpha}^{\dagger}\bm{\sigma}_{\alpha\beta}d_{\beta}$.
In addition a projection constraint ensures no-double occupancy of
the electron level $d_{\sigma}^{\dagger}$. The term proportional
to $J$ is the Heisenberg interaction between the dot and the spin,
while $J_{A}$ represents the process in which spin-conservation is
broken.

The QD with Hamiltonian Eq.~\ref{eq:Hdotspin} is tunnel-coupled
to the quantum spin Hall (QSH) edge \cite{Altshuler2013} through
a Hamiltonian \cite{Vaeyrynen2013,Vaeyrynen2014}
\begin{align}
H_{j-d} & =t\left[a_{\uparrow}^{\dagger}\left(x=0\right)d_{\uparrow}+a_{\downarrow}^{\dagger}\left(x=0\right)d_{\downarrow}+{\rm h.c.}\right]
\end{align}
 where $a_{\sigma}^{\dagger}\left(x=0\right)$ creates electrons on
the QSH edge. A time-reversal breaking impurity on the QSH edge is
expected to produce a $4\pi$-periodic Josephson effect because of
the flip of fermion parity with each $2\pi$ shift of the phase $\phi$.
While the spin in the QD acts as a magnetic impurity, as illustrated
in Fig.~\ref{fig:QDsys}(b), this only works in the case of odd fermion
parity of the JJ where by Kramers theorem the ground state is two-fold
degenerate. As will be shown, while the QD returns to odd FP state
each $4\pi$ period (as in the time-reversal breaking case), the spin
in the QD is flipped over each such period. This leads to the generic
$8\pi$-periodicity of the current as a function of phase.

To illustrate the mechanism in Fig.~\ref{fig:QDsys}(b) quantitatively,
we consider the limit of weak tunnel coupling $t$, the quantum dot
electron can only tunnel to a low-energy ABS in a Josephson junction
on the edge written as $\gamma^{\dagger}=\sum_{\sigma}\int dx\left(u_{\sigma}a_{\sigma}^{\dagger}+v_{\sigma}a_{\sigma}\right)$
with an energy $E\left(\phi\right)$, which depends on the phase difference
across the Josephson junction. The effective Hamiltonian of the edge
ABS is written as 
\begin{align}
 & H_{j}=E\left(\phi\right)(\gamma^{\dagger}\gamma-1/2). \label{eq:Hj2nd}
\end{align}
 The wavefunction of the ABS $\gamma$ and its energy are solved from
the BdG Hamiltonian of the JJ 
\begin{align}
{\cal H}_{{\rm BdG}} & =\tau_{z}\left(-iv_{F}s_{z}\partial_{x}-\mu\right)+\Delta\cos{\phi(x)}\tau_{x}+\Delta\sin{\phi(x)}\tau_{y}\label{eq:Hjunction}
\end{align}
 where $s$ and $\tau$ are Pauli matrices on spin and Nambu spaces,
respectively, and $\phi\left(x\right)=\phi\theta\left(x\right)$.
Since $\left[{\cal H}_{{\rm BdG}},s_{z}\right]=0$, the solutions
are labeled with the eigenvalues of $s_{z}$, where the solution with
$s_{z}=+1$ have $E\left(\phi\right)=-\Delta\cos\frac{\phi}{2}$,
$u_{\uparrow}\left(x=0\right)=v_{\downarrow}\left(x=0\right)^{*}=\sqrt{\frac{\sin\nicefrac{\phi}{2}}{2\xi}}e^{-i\phi/4}$
and $u_{\downarrow}=v_{\uparrow}=0$ in the interval $0\leq\phi\leq2\pi$,
while its particle-hole-conjugated partner is the other branch of
solution with $s_{z}=-1$. With this explicit form of quasiparticle
solution, we can therefore rewrite $H_{j-d}$ as 
\begin{equation}
H_{j-d}=t\left(u_{\uparrow}d_{\uparrow}^{\dagger}\gamma+v_{\downarrow}\gamma d_{\downarrow}+{\rm h.c.}\right),\label{eq:Hjdqp}
\end{equation}
 where we have dropped the position arguments of $u_{\uparrow}$ and
$v_{\downarrow}$.

The simplified form of the effective coupling allows us to describe
the cycle of the QD shown in Fig.~\ref{fig:QDsys}(b) as the phase
$\phi$ is varied. During each cycle of advancing $\phi$ by $2\pi$
forward, a quasiparticle is {}``pumped'' from the bulk occupied
states towards the conduction states through the edge states. The
excitation of a bulk conduction band electron must be avoided to prevent
dissipation. This can be accomplished by adding a spin-up electron
or removing a spin-down electron from the QD (and releasing a Cooper
pair). Starting with the state $\left|\downarrow\right\rangle $ at
$\phi=0$, i.e. with the dot empty and localized spin at $S_{z}=-1$.
During an increment of $2\pi$ of $\phi$, a spin-up electron is added
to the dot from the bulk, which due to the hybridization term Eq.~(\ref{eq:Hdotspin})
forms a singlet with the localized spin $\left(\frac{d_{\uparrow}^{\dagger}\left|\uparrow\right\rangle -d_{\downarrow}^{\dagger}\left|\downarrow\right\rangle }{\sqrt{2}}\right)$.
In the next cycle a spin-down electron is removed from the dot, leaving
the localized spin at $S_{z}=+1$ $\left(\left|\uparrow\right\rangle \right)$.
The electron leaving the dot can combine with the next electron coming
from the bulk and exit as a Cooper pair. The next two cycles proceed
similarly, with $J_{A}$ breaking spin conservation to result in a
triplet state after the third cycle $\left(\frac{d_{\uparrow}^{\dagger}\left|\uparrow\right\rangle -d_{\downarrow}^{\dagger}\left|\downarrow\right\rangle }{\sqrt{2}}\right)$
and returning to the original state $\left|\downarrow\right\rangle $
at $\phi=8\pi$.

\begin{figure}
\begin{centering}
\includegraphics[width=1\columnwidth]{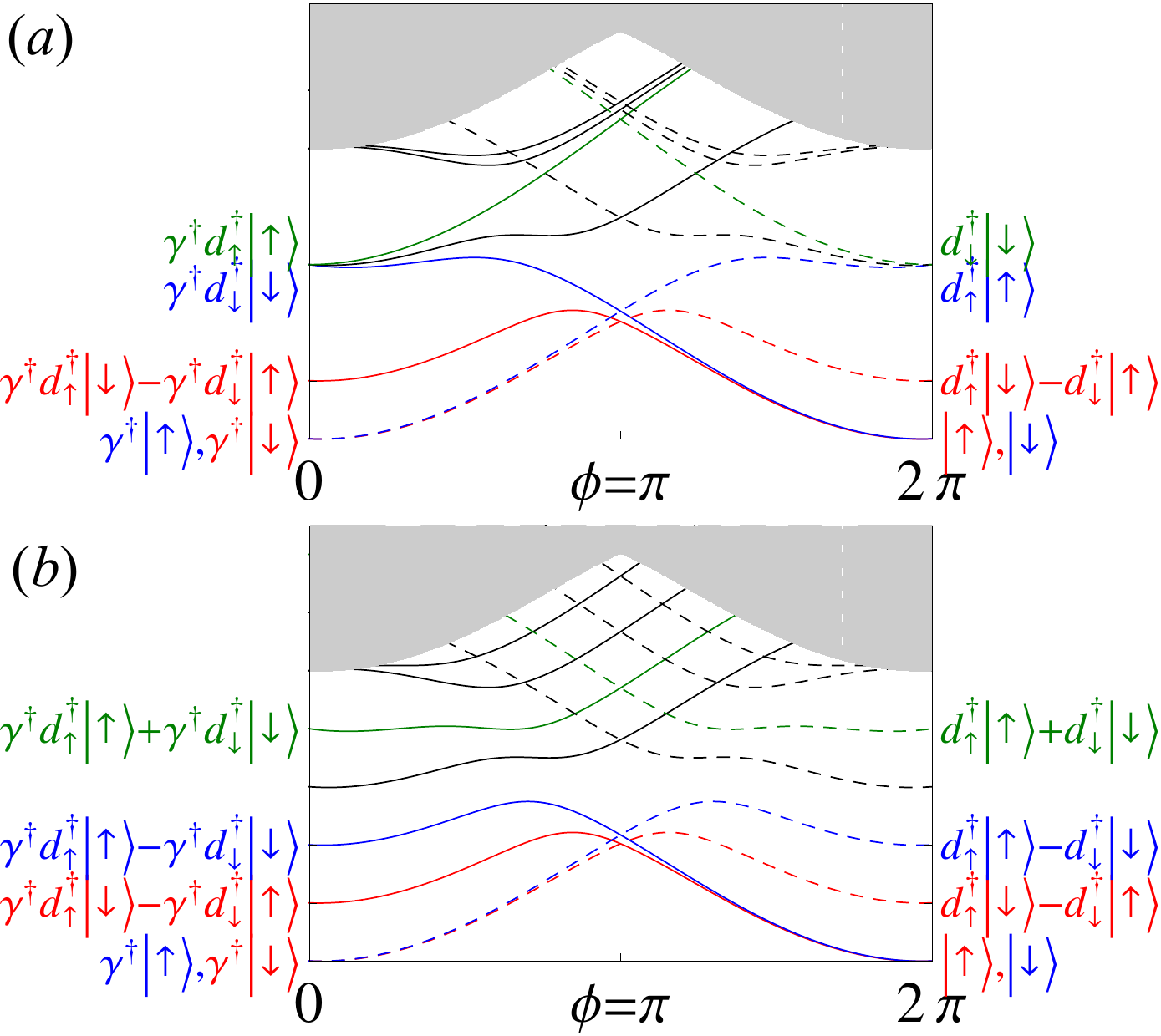} 
\par\end{centering}

\caption{The (many-electron) energy spectrum of a coupled
SC junction/QD/spin system [Eqs.~(\ref{eq:Hdotspin})-(\ref{eq:Hj2nd})] as
a function of superconducting phase $\phi$ with $J=0.2\Delta$ and
$\varepsilon_{d}=0.5\Delta$ and (a) $J_{A}=0$ (spin-conserving case)
and (b) $J_{A}=J$ (spin-anisotropic case). The solid lines represent
odd total fermion parity (Kramer's degenerate) states and the dashed
lines represent even total fermion parity states. The fermion parity
changes with each $2\pi$ cycle of $\phi$ and the color represents
the curves corresponding to various eigenstates. Following the spectrum
in (a) we find that the ground state necessarily couples to the bulk
states. In the spin-conservation breaking in (b), the spectrum is
isolated from the bulk states and $8\pi$ periodic. \label{fig:QDen}}
\end{figure}

The above process can be put on quantitative footing by projecting
the Hamiltonian into the low-energy Hilbert space and solving for
the energy-phase relation (E$\Phi$R), as detailed in Appendix A.
Typical results with $J_{A}$ being zero or non-zero are
shown respectively in Fig.~\ref{fig:QDen}(a) and \ref{fig:QDen}(b).
In the absence the term proportional to $J_{A}$, the full $8\pi$-cycle
could not be completed non-dissipatively because the dot-spin triplet
state could not be formed. Fig.~\ref{fig:QDen}(a) illustrates that
the state is eventually driven into continuum, thereby dissipating
away the Josephson current. Alternatively, dissipation of the excess
energy into a phonon might lead to a dissipative $4\pi-$periodic
process. With non-zero $J_{A}$ {[}Fig.~\ref{fig:QDen}(b){]}, a
full cycle of states fully gapped from other excited states can be
obtained. We also note that $J_{A}$ breaks spin conservation along
$z$-direction. The connection of the absence of spin conservation
with the prevention of dissipation will be elaborated below.

\section{Conditions for dissipationless Josephson effect}

\begin{figure}
\begin{centering}
\includegraphics[width=1\columnwidth]{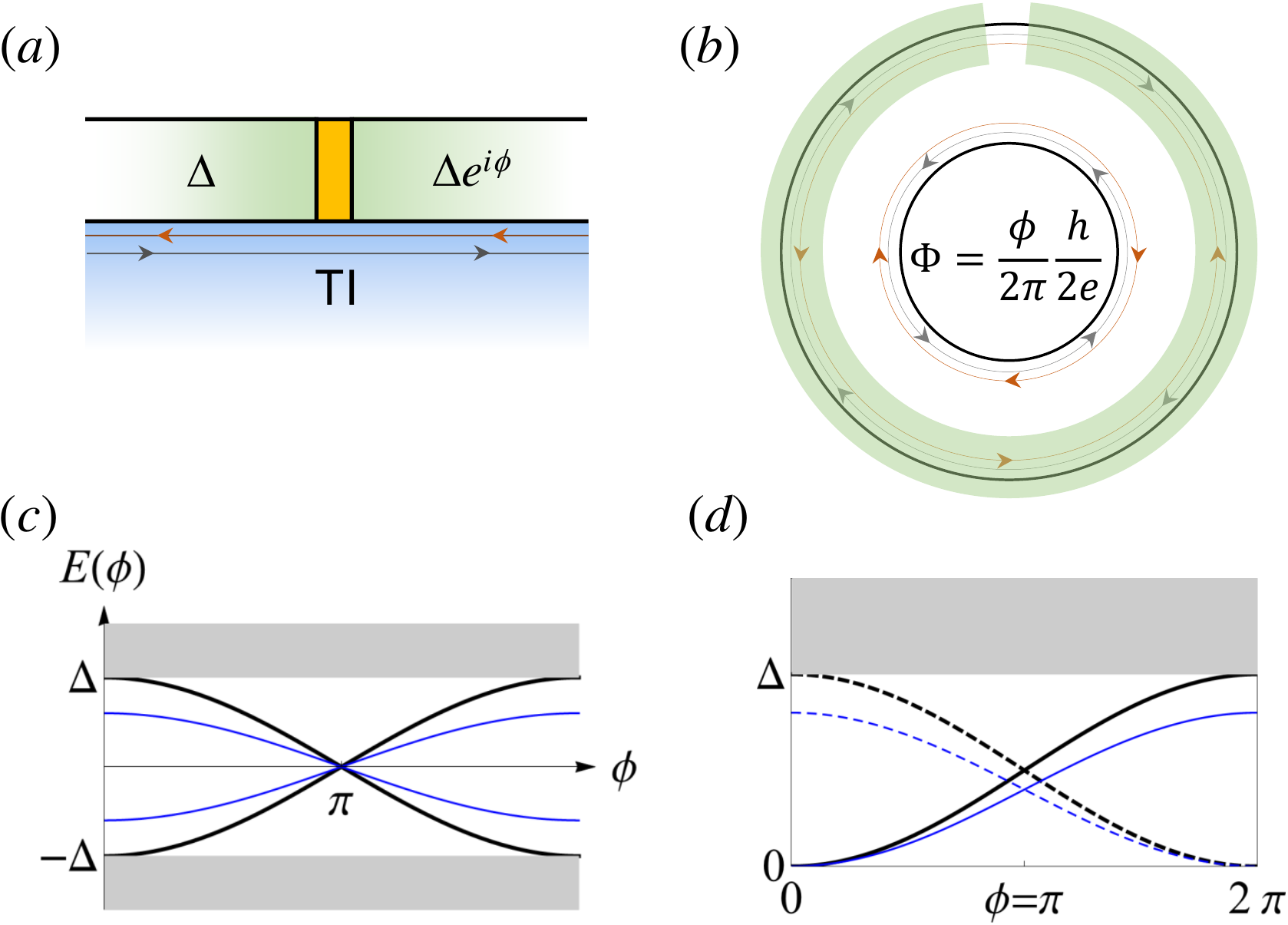} 
\par\end{centering}

\caption{(a) A short Josephson junction on a one-dimensional TI edge. (b) Equivalent
construction where now the phase difference is controlled by threading
a flux through the center of the Corbino disk. (c) The single-particle
Andreev spectrum for the junction with (blue thin lines) or without
(black thick lines) the FI element. (d) The many-body E$\Phi$R with
(blue thin lines) or without (black thick lines) the FI element, where
solid (dotted) lines indicate states with even (odd) parity.\label{fig:JJ}}
\end{figure}

We now
discuss in general the necessary conditions to realize a topological
TRS Josephson junction without dissipation. We first review how this
is accomplished in the case where time-reversal symmetry is broken
by an FI element in a topological junction with only one ABS (see
Fig.~\ref{fig:JJ}a). The E$\Phi$R is shown in Fig.~\ref{fig:JJ}(c).
At $\phi\neq2\pi p$, where $p$ is an integer, a particle-hole pair
of ABS is present in the junction. Without the FI which breaks TRS,
these states are required to join the continuum modes ($\left|E\right|=\Delta$)
at the time-reversal-invariant points $\phi=2\pi p$ in order to satisfy
the Kramers theorem (black lines). This requirement corresponds (see
Fig.~\ref{fig:JJ}(d)) to the adiabatic change of the phase $\phi$
connecting the ground state at $\phi=0$ with the continuum at $\phi=2\pi$.
This creates quasiparticle excitations in the bulk that lead to dissipation
in the Josephson junction. On the other hand, with the FI that breaks
TRS even at $\phi=2\pi p$, the ABSs remain disconnected from the
continuum modes {[}blue lines in Fig.~\ref{fig:JJ}(c){]}, and thus
the many-body state remains gapped from the continuum {[}blue lines
in Fig.~\ref{fig:JJ}(d){]}. The full $4\pi$-cycle of Josephson
current could then be completed without dissipation if the temperature
and rate of change of $\phi$ are low enough.

\begin{figure}
\begin{centering}
\includegraphics[width=1\columnwidth]{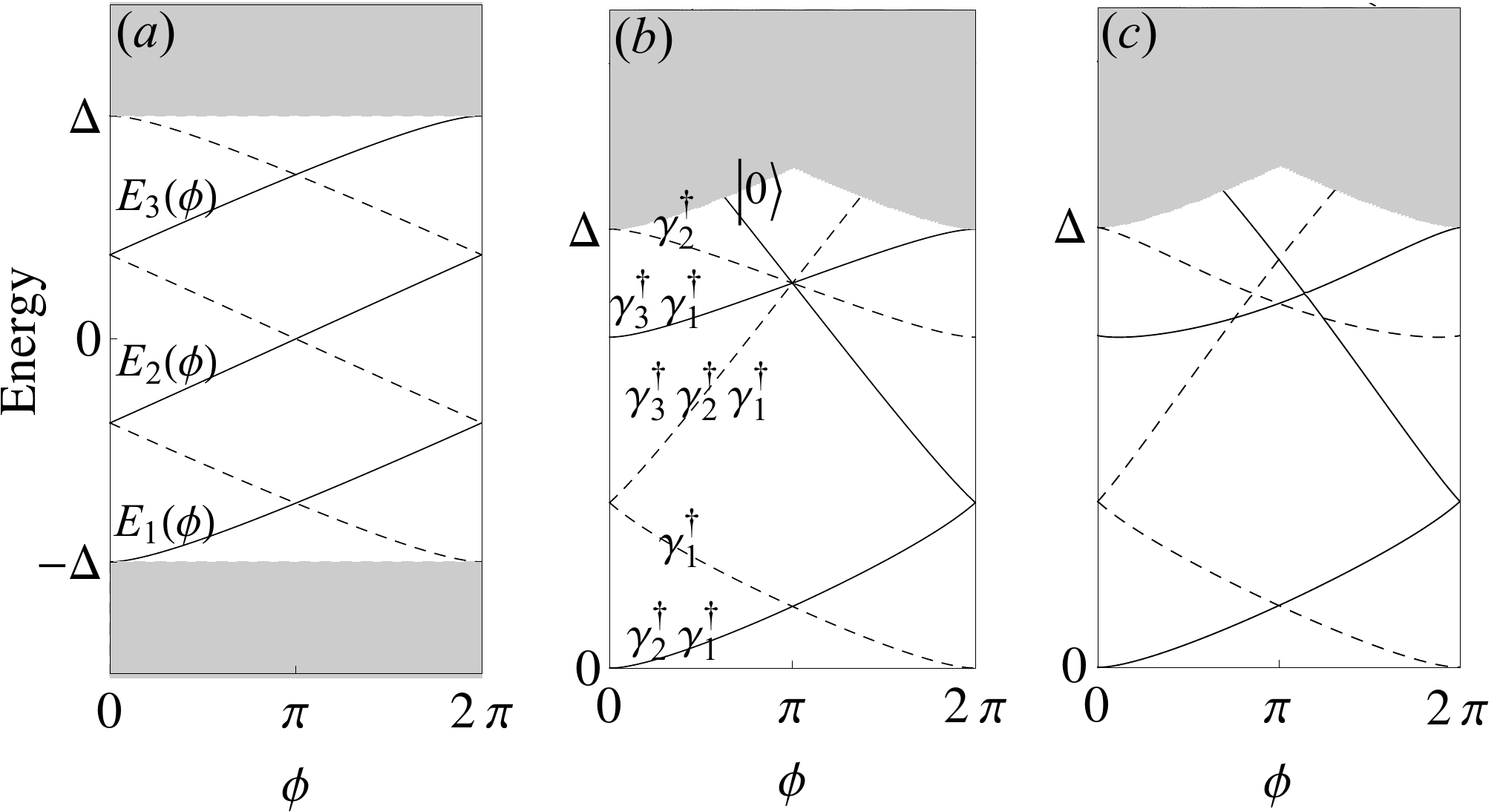} 
\par\end{centering}

\caption{\label{fig:LongInt}(a) E$\Phi$R for a junction with length $L=\pi v_{F}/\Delta_{0}$.
The solid (dotted) lines are solutions with eigenvalues $s_{z}=\pm1$
(b) Many-body spectrum for the same non-interacting junction. Black
(dotted) lines are states with even (odd) parity. The labels indicate
which of the ABSs are occupied. (c) Many-body spectrum for a junction
with $s_{z}$ conserving interactions. Detailed calculations are presented
in Appendix B.}
\end{figure}

As seen in Fig.~\ref{fig:LongInt}(a), the ABS levels in the non-interacting
case connect the valence bands to the conduction bands~ \cite{Fu2009}.
This is a necessary consequence of Kramer's degeneracy and the time-reversal
(i.e.. $\phi\rightarrow-\phi$) properties of the eigenvalues shown
in Fig.~\ref{fig:LongInt}(a). This leads us to conclude that interactions
are necessary to avoid dissipation in TRS topological Josephson junctions.
Next we argue that spin-conservation breaking is crucial to avoid
dissipation in the TI Josephson junction. To understand this, consider
a Josephson junction built from a Corbino geometry (shown at Fig.~\ref{fig:JJ}b),
in which the phase difference $\phi$ between the two sides of the
junction is controlled by threading a flux $\Phi_{{\rm flux}}=\frac{\phi}{2\pi}\Phi_{0}$
through the center of the setup, where $\Phi_{0}=\frac{h}{2e}$ is
the SC flux quantum. For the sake of the argument, we first take out
the SC and FI elements, leaving behind a Corbino disk made of TI.
It is known that if spin is conserved along the polarization axis
of the TI, the system exhibits quantized spin Hall conductance, and
by Laughlin's argument \cite{Laughlin1981} threading a flux quantum
has the effect of pumping a pair of spins with $s_{z}=\pm1$ to opposite
edges. This effect still holds with the introduction of the SC, since
$s$-wave superconductivity preserves $s_{z}$ conservation. In this
scenario, the portion of TI edge not in contact with the SC can provide
a finite number of ABSs (say, $n_{A}$) to accommodate the pumped
spins. After threading $n_{A}$ SC flux quanta (corresponding to an
incrementation of $\phi$ by $2\pi n_{A}$), the ABSs fail to accommodate
\emph{all} of the pumped spins, leading to the occupation of the other
continuum modes on the edges, which corresponds to a dissipation in
the Josephson junction. The way to avoid this is to break $s_{z}$-conservation
which would then destroy the quantized spin Hall conductance of the
system. Threading a flux quantum would then flip the Fermion parities
(FPs) of the two edges \cite{Fu2009}.

Based on these two observations, we expect that a dissipationless
Josephson junction in a TI requires an interaction term that breaks
$s_{z}$-conservation to a junction. This agrees with what we found
in the Josephson junction coupled with a quantum dot as described
above, where the term $J_{A}$ provides the necessary breaking of
spin conservation, without which dissipation could not be avoided.

\section{Role of spin-anisotropy}

To gain further insight into the
necessity of breaking spin conservation to obtain a dissipationless
TRS topological junction, we look into another example, first studied
in Ref.~\cite{Zhang2014}. Consider an SC-N-SC junction on a TI edge
where the N portion is long enough with multiple ABSs present. The
Hamiltonian is almost identical to our previous example ( Eq.~(\ref{eq:Hjunction}))
except that $\Delta$ is chosen to represent a long junction as $\Delta\left(x\right)=\Delta_{0}\theta\left(\left|x\right|-\frac{L}{2}\right)$.
When $L=\pi v_{F}/\Delta$, three ABSs are present for all values
of $\phi$, and the single-particle and many-body E$\Phi$R are shown
in Fig.~\ref{fig:LongInt}(a) and \ref{fig:LongInt}(b), respectively.
The key feature in Fig.~\ref{fig:LongInt}(b), which is needed to
understand the Josephson behavior, is the four-fold degeneracy at
$\phi=\pi$. As discussed in Ref.~\cite{Zhang2014}, splitting this
degeneracy by Coulomb interactions into to two two-fold Kramer's degenerate
crossings. However, as seen in Fig.~\ref{fig:LongInt}(c), the states
still continue to reach the continuum in the absence of spin conservation
breaking interactions. This forbids a dissipationless ac Josephson
effect in this case. To understand this, we note that the four states
have different number of quasiparticles: $\left\{ \left|0\right\rangle ,\,\gamma_{2}^{\dagger}\left|0\right\rangle ,\,\gamma_{3}^{\dagger}\gamma_{1}^{\dagger}\left|0\right\rangle ,\,\gamma_{3}^{\dagger}\gamma_{2}^{\dagger}\gamma_{1}^{\dagger}\left|0\right\rangle \right\} $,
and therefore have different values of $s_{z}$. Once the four-fold
degeneracy is lifted we are left with two-fold degeneracies at level
crossings. These crossings are however between states of different
$s_{z}$. In accordance with our general arguments presented above,
we find that $s_{z}$-conserving interaction terms (e.g. $\int dx\left(a_{\uparrow}^{\dagger}a_{\uparrow}+a_{\downarrow}^{\dagger}a_{\downarrow}\right)^{2}$)
in the Hamiltonian cannot split these crossings and the ground state
necessarily reaches the continuum simply by adiabatic evolution. Adding
$s_{z}$-breaking interactions like $\int dx\left(a_{\uparrow}^{\dagger}a_{\uparrow}a_{\downarrow}^{\dagger}\partial_{x}a_{\uparrow}-a_{\downarrow}^{\dagger}a_{\downarrow}a_{\uparrow}^{\dagger}\partial_{x}a_{\downarrow}\right)+{\rm h.c.}$
, which were assumed to be comparable to the Coulomb interactions
in Ref.~\cite{Zhang2014} are required to split the crossings to
avoid dissipation.

\section{General Theorem for $8\pi$-periodicity} 

The two examples
of dissipationless TRS topological junctions above both exhibit $8\pi$-periodicities.
We now argue that this is directly a consequence of TRS present in
the junction. To see this, consider a TRS junction described locally
by the Hamiltonian $H\left(\phi\right)$ where $\phi$ is the phase
difference of the constituent SCs. This Hamiltonian satisfies 
\begin{equation}
H\left(\phi\right)=H\left(\phi+2\pi\right)=\Theta H\left(-\phi\right)\Theta^{-1},\label{eq:H}
\end{equation}
 where the first condition follows from the $2\pi$-periodicity of
the phase of SC. The second condition, where $\Theta$ is the time-reversal
operator, follows from the time-reversal of the magnetic flux that
creates the superconducting phase $\phi$. Let us suppose that $\left|n\right\rangle _{\phi}$
is the $n^{{\rm th}}$ excited many-body state at phase $\phi$ satisfying
$H\left(\phi\right)\left|n\right\rangle _{\phi}=E_{n}\left(\phi\right)\left|n\right\rangle _{\phi}$
and $FP\left|n\right\rangle _{\phi}=\lambda_{n}\left(\phi\right)\left|n\right\rangle _{\phi}$,
where $FP$ is the Fermion parity operator and $E_{n}$ and $\lambda_{n}$
are respectively its energy and FP eigenvalues with $\lambda_{n}\left(\phi\right)=\pm1$
when there are even/odd number of fermions in the system, respectively.
It then follows from Eq.~(\ref{eq:H}) that 
\begin{equation}
E_{n}\left(\phi\right)=E_{n}\left(\phi+2\pi\right)=E_{n}\left(-\phi\right).\label{eq:E}
\end{equation}
 Finally, we stipulate that the FP must switch as $\phi$ is advanced
by $2\pi$, i.e., 
\begin{equation}
\lambda_{n}\left(\phi+2\pi\right)=-\lambda_{n}\left(\phi\right),\label{eq:Parity}
\end{equation}
 which describes the topological property of the TI. More specifically,
threading a flux through the TI Corbino disk (see Fig.~\ref{fig:JJ}b)
changes the $Z_{2}$ {}``time-reversal polarization'' \cite{Fu2006},
which can be identified with the FP of each of the edges \cite{Fu2009}.
We remark that this property is not captured by the local Hamiltonian
Eq.~(\ref{eq:H}) of the junction.

\begin{figure}
\begin{centering}
\includegraphics[width=0.8\columnwidth]{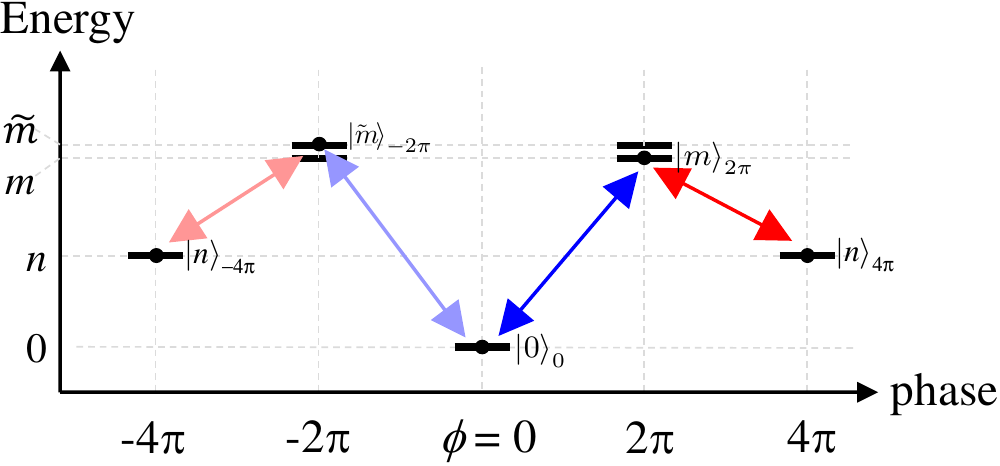} 
\par\end{centering}

\caption{\label{fig:proofe} The path of an adiabatically-followed even-parity
state. Blue (red) and light blue (light red) lines are time-reversed
paths of each other. The crucial point is $n\neq0$, as otherwise
$\left|0\right\rangle _{0}$ would join to $\left|\tilde{m}\right\rangle _{2\pi}$
and $\left|m\right\rangle _{2\pi}$ simultaneously.}
\end{figure}

The proof of $8\pi$-periodicity for a state with even FP is illustrated
in Fig.~\ref{fig:proofe}, and the mathematically rigorous proof
is given in the Appendix C. Let us start with a non-degenerate
state $\left|0\right\rangle _{0}$. Tuning $\phi$ forward and backward
by $2\pi$ reaches the degenerate states $\left|m\right\rangle _{2\pi}$
and $\left|\tilde{m}\right\rangle _{-2\pi}$ (recalling that by Eq.~(\ref{eq:Parity})
$\left|m\right\rangle _{0}$ and $\left|\tilde{m}\right\rangle _{0}$
are odd-parity states and is a Kramers pair at $\phi=0$). Further
increasing or decreasing $\phi$ by $2\pi$, the state $\left|n\right\rangle _{\pm4\pi}$
is reached. From Eq.~(\ref{eq:E}) we know that $\left|n\right\rangle _{4\pi}\neq\left|0\right\rangle _{0}$
because $\left|0\right\rangle _{0}$ cannot be adiabatically connected
both $\left|m\right\rangle _{2\pi}$ and $\left|\tilde{m}\right\rangle _{2\pi}$
at the same time (compare, e.g., light red and deep blue lines in
Fig.~\ref{fig:proofe}). The proof for a state with odd FP proceeds
in a similar way and is discussed in the appendix.

In summary, interactions and spin-conservation breaking are the two
key ingredients that are required to permit a dissipationless ac Josephson
effect in a TRS topological Josephson junction. In this paper, we
have given a general proof that the E$\Phi$R of such a dissipationless
TRS topological Josephson junction is $8\pi$-periodic in $\phi$.
The $8\pi$ periodicity arises from the combination of the flip of
FP and spin over each $2\pi$ period. We have shown that these ingredients
can be incorporated naturally in a model of a quantum dot coupled
to a Josephson junction on a TI edge.

This work is supported by JQI-NSF-PFC and the University of Maryland
startup grant. HH acknowledge support from AFOSR (FA9550-15-1-0445)
and ARO (W911NF-16-1-0182) during the final stage of this work. We
acknowledge the discussion with Carlo Beenakker.

Note added-- During the preparation of the manuscript we became aware
of a related recent-published work by Peng et al. \cite{Peng2016}.

\appendix
\begin{widetext}

\section{Low-energy Hamiltonian for the Junction-Dot System}

Since the Hamiltonian (\ref{eq:Hdotspin}) conserves the parity of
electron number, we expand it in odd and even parity subspaces. We
also take the limit $U\rightarrow\infty$, which projects out the
states where the quantum dot is doubly-occupied. The basis states
for the odd-parity subspace are $\left\{ \gamma^{\dagger}\left|\uparrow\right\rangle ,\gamma^{\dagger}\left|\downarrow\right\rangle ,d_{\uparrow}^{\dagger}\left|\downarrow\right\rangle ,d_{\downarrow}^{\dagger}\left|\uparrow\right\rangle ,d_{\uparrow}^{\dagger}\left|\uparrow\right\rangle ,d_{\downarrow}^{\dagger}\left|\downarrow\right\rangle \right\} $,
where $\left|\nicefrac{\uparrow}{\downarrow}\right\rangle $ satisfies
$S^{z}\left|\nicefrac{\uparrow}{\downarrow}\right\rangle =\pm\left|\nicefrac{\uparrow}{\downarrow}\right\rangle $
and $\gamma\left|\nicefrac{\uparrow}{\downarrow}\right\rangle =d_{\sigma}\left|\nicefrac{\uparrow}{\downarrow}\right\rangle =0$.
The Hamiltonian in this basis is 
\begin{equation}
H^{\left(o\right)}=\left(\begin{array}{cccccc}
\frac{E\left(\phi\right)}{2} & 0 & 0 & 0 & tu_{\uparrow}^{*} & 0\\
0 & \frac{E\left(\phi\right)}{2} & tu_{\uparrow}^{*} & 0 & 0 & 0\\
0 & tu_{\uparrow} & -\frac{E\left(\phi\right)}{2}-\frac{J}{2}+\varepsilon_{d} & J & 0 & 0\\
0 & 0 & J & -\frac{E\left(\phi\right)}{2}-\frac{J}{2}+\varepsilon_{d} & 0 & 0\\
tu_{\uparrow} & 0 & 0 & 0 & -\frac{E\left(\phi\right)}{2}+\frac{J}{2}+\varepsilon_{d} & J_{A}\\
0 & 0 & 0 & 0 & J_{A} & -\frac{E\left(\phi\right)}{2}+\frac{J}{2}+\varepsilon_{d}
\end{array}\right),
\end{equation}
 while in the even subspace with basis $\left\{ \left|\uparrow\right\rangle ,\left|\downarrow\right\rangle ,\gamma^{\dagger}d_{\uparrow}^{\dagger}\left|\downarrow\right\rangle ,\gamma^{\dagger}d_{\downarrow}^{\dagger}\left|\uparrow\right\rangle ,\gamma^{\dagger}d_{\uparrow}^{\dagger}\left|\uparrow\right\rangle ,\gamma^{\dagger}d_{\downarrow}^{\dagger}\left|\downarrow\right\rangle \right\} $,
the Hamiltonian is expanded as 
\begin{equation}
H^{\left(e\right)}=\left(\begin{array}{cccccc}
-\frac{E\left(\phi\right)}{2} & 0 & 0 & -tv_{\downarrow} & 0 & 0\\
0 & -\frac{E\left(\phi\right)}{2} & 0 & 0 & 0 & -tv_{\downarrow}\\
0 & 0 & \frac{E\left(\phi\right)}{2}-\frac{J}{2}+\varepsilon_{d} & J & 0 & 0\\
-tv_{\downarrow}^{*} & 0 & J & \frac{E\left(\phi\right)}{2}-\frac{J}{2}+\varepsilon_{d} & 0 & 0\\
0 & 0 & 0 & 0 & \frac{E\left(\phi\right)}{2}+\frac{J}{2}+\varepsilon_{d} & J_{A}\\
0 & -tv_{\downarrow}^{*} & 0 & 0 & J_{A} & \frac{E\left(\phi\right)}{2}+\frac{J}{2}+\varepsilon_{d}
\end{array}\right).
\end{equation}

Finally, we note that at $\phi=2n\pi$, $v_{\downarrow}=u_{\uparrow}=0$
which enables us to reach the simple forms of eigenstates shown in
Fig.~\ref{fig:QDsys}(c,d)

\section{Bogoliubov-de Gennes Solution for a Long Topological Junction}

Since $s_{z}$ commutes with ${\cal H}_{{\rm BdG}}$, the solutions
to ${\cal H}_{{\rm BdG}}\psi_{n}=E_{n}\psi_{n}$ are labeled by the
{}``spin'' index $s_{z}=\pm1$. The $s_{z}=+1$ solutions, denoted
by $\psi_{n}^{(+)}$, have $u_{n\downarrow}^{(+)}=v_{n\uparrow}^{(+)}=0$
and \begin{subequations}\label{eq:uv-1} 
\begin{align}
u_{n\uparrow}^{(+)} & ={\cal A}_{n}e^{{\rm sgn}\left(x\right)i\theta_{E_{n}}/2}e^{i\frac{\phi\left(x\right)}{2}+i\bar{\mu}\bar{x}-\sqrt{1-\bar{E}_{n}^{2}}\left|\left|\bar{x}\right|-\frac{\bar{L}}{2}\right|}\\
v_{n\downarrow}^{(+)} & ={\cal A}_{n}e^{-{\rm sgn}\left(x\right)i\theta_{E_{n}}/2}e^{-i\frac{\phi\left(x\right)}{2}+i\bar{\mu}\bar{x}-\sqrt{1-\bar{E}_{n}^{2}}\left|\left|\bar{x}\right|-\frac{\bar{L}}{2}\right|}
\end{align}
 for $\left|x\right|>\frac{L}{2}$, and 
\begin{align}
u_{n\uparrow}^{(+)} & ={\cal A}_{n}e^{-i\theta_{E_{n}}/2}e^{i\bar{\mu}\bar{x}+i\bar{E}_{n}\left(\bar{x}+\frac{\bar{L}}{2}\right)}\\
v_{n\downarrow}^{(+)} & ={\cal A}_{n}e^{i\theta_{E_{n}}/2}e^{i\bar{\mu}\bar{x}-i\bar{E}_{n}\left(\bar{x}+\frac{\bar{L}}{2}\right)}
\end{align}
 \end{subequations}otherwise. Here the normalization factor is ${\cal A}_{n}=\left(2L+2\xi/\sqrt{1-\bar{E}_{n}^{2}}\right)^{-1/2}$,
and $\bar{E}=\frac{E}{\Delta_{0}}$, $\bar{\mu}=\frac{\mu}{\Delta_{0}}$,
$\bar{x}=\frac{x}{\xi}=\frac{x}{v_{F}/\Delta_{0}}$, $\bar{L}=\frac{L}{\xi}$,
$e^{\pm i\theta_{E_{n}}}=\bar{E}_{n}\pm i\sqrt{1-\bar{E}_{n}^{2}}$,
where $\bar{E}_{n}$ satisfies 
\begin{equation}
\sqrt{1-\left(\frac{E_{n}}{\Delta_{0}}\right)^{2}}\cos\left(\frac{E_{n}L}{v_{F}}-\frac{\phi}{2}\right)-\frac{E_{n}}{\Delta_{0}}\sin\left(\frac{E_{n}L}{v_{F}}-\frac{\phi}{2}\right)=0.\label{eq:En}
\end{equation}

The $s_{z}=-1$ solutions $\psi_{n}^{(-)}$ with energy $-E_{n}$
are related to $\psi_{n}^{(+)}$ by particle-hole conjugation, $\psi_{n}^{(-)}=\Xi\psi_{n}^{(+)}$,
where $\Xi=s_{y}\tau_{y}K$. With the solutions to ${\cal H}_{{\rm BdG}}$
we can expand the Hamiltonian in quasiparticle operators as 
\begin{equation}
H_{0}\left(\phi\right)=\sum_{n}E_{n}\left(\phi\right)\left(\gamma_{n}^{\dagger}\gamma_{n}-\frac{1}{2}\right)
\end{equation}
 where $E_{n}$ are determined from Eq.~(\ref{eq:En}) and $\gamma_{n}^{\dagger}=\sum_{\sigma=\uparrow/\downarrow}\int dx\left(u_{n\sigma}^{(+)}a_{\sigma}^{\dagger}+v_{n\sigma}^{(+)}a_{\sigma}\right)$.
Since only the branch of solutions with $s_{z}=+1$ are summed, the
number of quasiparticles in a many-body state coincide with the value
of $s_{z}$ for that state.

\section{Mathematically Rigorous Proof of 8$\pi$ -Periodicity}

Let $U_{\phi_{2},\phi_{1}}=e^{-i\int_{\phi_{1}}^{\phi_{2}}H\left(\phi\right)d\phi}$
be the operator that adiabatically changes the phase from $\phi_{1}$
to $\phi_{2}$. The conditions\begin{subequations}\label{eqs:U}
\begin{align}
\left|m\right\rangle _{\phi_{2}} & =U_{\phi_{2},\phi_{1}}\left|n\right\rangle _{\phi_{1}}\Leftrightarrow\left|n\right\rangle _{\phi_{1}}=U_{\phi_{1},\phi_{2}}\left|m\right\rangle _{\phi_{2}}\\
 & \Leftrightarrow\left|m\right\rangle _{2p\pi+\phi_{2}}=U_{2p\pi+\phi_{2},2p\pi+\phi_{1}}\left|n\right\rangle _{2p\pi+\phi_{1}},\label{eq:Uperiodic}
\end{align}
 \end{subequations}for all integers $p$, follows directly from the
unitarity of $U$ and Eq.~(\ref{eq:E}).

We first consider the case where the ground state at $\phi=0$, $\left|0\right\rangle _{0}$,
has even FP, i.e., $\lambda_{0}\left(0\right)=1$ (Fig.~\ref{fig:proofe}).
Starting with $\left|0\right\rangle _{0}$, as the phase is adiabatically
tuned to $2\pi$ and $4\pi$, the state is brought to the $m^{{\rm th}}$
and $n^{{\rm th}}$ excited states respectively, i.e.\begin{subequations}\label{eqs:evenPosPhi}
\begin{align}
U_{2\pi,0}\left|0\right\rangle _{0} & =\left|m\right\rangle _{2\pi},\label{eq:evenPosPhia}\\
U_{4\pi,2\pi}\left|m\right\rangle _{2\pi} & =\left|n\right\rangle _{4\pi}.
\end{align}
 \end{subequations}Now we know $m\neq0$ because from Eqs.~(\ref{eq:Parity})
$\lambda_{0}(2\pi)=-\lambda_{0}(0)=-1$, i.e. the FP of the ground
state at $\phi=2\pi$ is odd. The state that $\left|0\right\rangle _{0}$
transforms into must be even in FP, i.e. $\lambda_{m}\left(2\pi\right)=\lambda_{0}\left(0\right)=1$,
and from Eq.~(\ref{eq:Parity}) we have $\lambda_{m}\left(0\right)=-1$,
i.e. $\left|m\right\rangle _{0}$ is odd in FP. Since $H\left(0\right)$
is TRS, Kramers theorem guarantees that there is an orthogonal state
$\left|\tilde{m}\right\rangle _{0}=\Theta\left|m\right\rangle _{0}$
with the same energy: $E_{\tilde{m}}\left(0\right)=E_{m}\left(0\right)$.
Apply $\Theta$ on Eqs.~(\ref{eqs:evenPosPhi}) we have\begin{subequations}
\begin{align}
U_{-2\pi,0}\left|0\right\rangle _{0} & =\left|\tilde{m}\right\rangle _{-2\pi},\\
U_{-4\pi,-2\pi}\left|\tilde{m}\right\rangle _{-2\pi} & =\left|n\right\rangle _{-4\pi}.\label{eq:evenNegPhib}
\end{align}
 \end{subequations}

We have thus obtained an energy-phase relation (E$\Phi$R) schematically
shown in Fig.~\ref{fig:proofe} which is $8\pi$-periodic \emph{if}
$n\neq0$ (and assuming no accidental degeneracy $E_{n}\neq E_{0}$).
To establish this, we use Eq.~(\ref{eqs:U}) to derive from Eq.~(\ref{eq:evenNegPhib})
\begin{equation}
U_{2\pi,0}\left|n\right\rangle _{0}=\left|\tilde{m}\right\rangle _{2\pi},\label{eq:evenProof}
\end{equation}
 which can be compared with Eq.~(\ref{eq:evenPosPhia}). Since $U_{0,2\pi}\left|m\right\rangle _{2\pi}=\left|0\right\rangle _{0}$,
we know $U_{0,2\pi}\left|\tilde{m}\right\rangle _{2\pi}\neq\left|0\right\rangle _{0}$
(as $\left|0\right\rangle _{0}$ cannot be adiabatically connected
to two states at $\phi=2\pi$) and from Eq.~(\ref{eq:evenProof})
this means $n\neq0$. The $8\pi$-periodic E$\Phi$R is therefore
established.

\begin{figure}
\begin{centering}
\includegraphics[width=0.5\columnwidth]{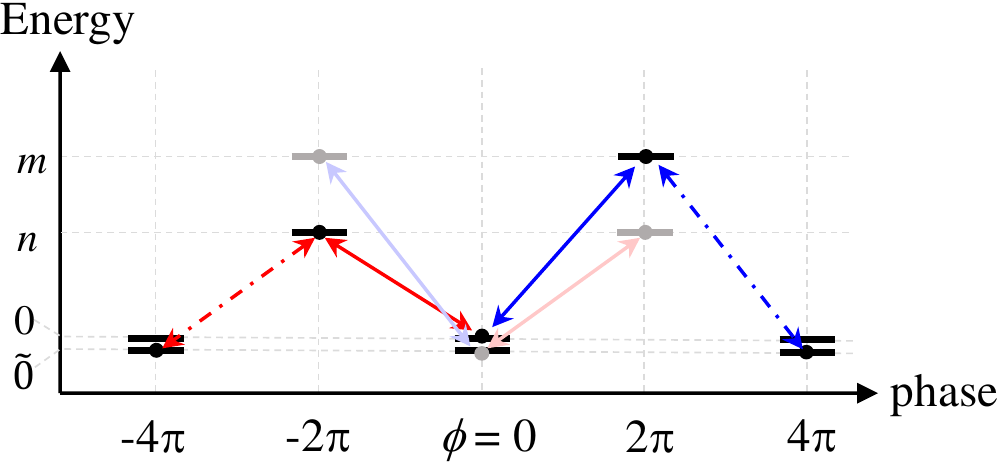} 
\par\end{centering}

\caption{\label{fig:proofo} For odd-parity state we define $m$ and $n$ as
the states that Kramers pair $\left\{ \left|0_{0}\right\rangle ,\left|\tilde{0}_{0}\right\rangle \right\} $
traverses as a flux quantum is inserted in either direction. Here
again deep-colored and light-colored lines are time-reversed paths
of each other. Use Eq.~(\ref{eq:E}) to shift the light blue and
light red lines by $4\pi$, we get the dashed path, completing the
full $8\pi$-periodic path that the state follows.}
\end{figure}

The case of $\lambda_{0}\left(0\right)=-1$ can be considered in a
similar fashion (Fig.~\ref{fig:proofo}). Let $\left|0\right\rangle _{0}$
and $\left|\tilde{0}\right\rangle _{0}$ be the Kramers pair of degenerate
ground states at $\phi=0$. Define $\left|m\right\rangle _{2\pi}$
and $\left|n\right\rangle _{-2\pi}$ be respectively the states that
$\left|0\right\rangle _{0}$ transforms into as $\phi$ is tuned from
$0$ to $\pm2\pi$, respectively. We have 
\begin{align}
U_{2\pi,0}\left|0\right\rangle _{0}=\left|m\right\rangle _{2\pi},\: & U_{-2\pi,0}\left|0\right\rangle _{0}=\left|n\right\rangle _{-2\pi},\label{eq:odd1}\\
U_{2\pi,4\pi}\left|\tilde{0}\right\rangle _{4\pi}=\left|m\right\rangle _{2\pi},\: & U_{-2\pi,-4\pi}\left|\tilde{0}\right\rangle _{-4\pi}=\left|n\right\rangle _{-2\pi},\label{eq:odd2}
\end{align}
 where the second line is obtained by applying $\Theta$ and Eq.~(\ref{eq:Uperiodic})
subsequently on the first line. Finally, we use Eq.~(\ref{eq:Uperiodic})
to derive, from the last relation above, 
\begin{equation}
U_{2\pi,0}\left|\tilde{0}\right\rangle _{0}=\left|n\right\rangle _{2\pi}.
\end{equation}
 Upon comparison with the first relation of Eq.~(\ref{eq:odd1}),
this shows that $n\neq m$ and hence the full $8\pi$-periodic cycle
of E$\Phi$R shown in Fig.~\ref{fig:proofo} (with $E_{n}\neq E_{m}$)
is established.\end{widetext}

\vfill{}

 \bibliographystyle{apsrev4-1}
\bibliography{TRIJosephson}

\end{document}